\documentstyle{rs}

\begin{document}

\title[Bose-Einstein Condensation Theory]{Theory of Bose-Einstein condensation for trapped atoms}

\author[N. P. Proukakis and K. Burnett]{Nick P. Proukakis and Keith Burnett}

\affiliation{Clarendon Laboratory, Department of Physics, University of Oxford, \\ Parks Road, Oxford OX1 3PU, U.K.}
\maketitle

\begin{center}
[ To appear in Phil. Trans. R. Soc. Lond. A 355 (1997). ]\\
\end{center}

\vspace{1.0cm}

\begin{abstract}

We outline the general features of the conventional mean field theory for the description of Bose-Einstein condensates at near zero temperatures. This approach, based on a phenomenological model, appears to give excellent agreement with experimental data. We argue, however, that such an approach is not rigorous and cannot contain the full effect of collisional dynamics due to the presence of the mean field. We thus discuss an alternative microscopic approach and explain, within our new formalism, the physical origin of these effects. Furthermore, we discuss the potential formulation of a consistent finite-temperature mean field theory, which we claim necessitates an analysis beyond the conventional treatment.

\end{abstract}

\section{Introduction}

It has been known for many years that the quantum nature of atomic motion can produce drastic effects on the properties of a gas. At sufficiently low temperatures, the atomic de Broglie waves of neighbouring atoms in an assembly start overlapping, giving rise to the quantum statistical effects that discriminate between fermions and bosons (depending on their atomic spin). The Pauli exclusion principle prohibits any two fermions from occupying exactly the same quantum state. On the other hand, bosons are not limited in this way, and an arbitrary number of bosons can occupy the same quantum state. In fact, when the de Broglie waves start to overlap, the process of Bose-Einstein Condensation (BEC) occurs, in which bosonic atoms produce a macroscopic occupancy of a single state, usually the ground state of the container they are in. An additional and  fundamental characteristic of BEC is the acquisition of a well-defined phase of the whole condensate, in much the same way magnetisation occurs in the paramagnetic to ferromagnetic transition\footnote{It is worth pointing out that, despite the Pauli exclusion principle, fermions can form stable pairs, known as Cooper pairs, whose cooperative effect gives rise to the well-known phenomenon of superconductivity (flow of electric charge in the absence of resistance).}.

The process of condensation is driven by the quantum statistics as follows (see e.g. Burnett 1996): each collision between atoms results in scattering into any pair of states allowed by energy and momentum conservation. When, however, one takes quantum statistics into account, one finds that scattering is enhanced into those states that already have some atoms in them. This means that once a number of atoms has `condensed' into the ground state, collisions between the other atoms in the assembly will tend to increase the number of atoms in that state. This will lead to BEC, provided other processes that produce loss of atoms from the condensate don't dominate. The precise details of the onset of BEC are still a matter of research. What is, however, undisputed is the fact that the phenomenon occurs when the assembly is sufficiently cold and dense for the atomic waves to overlap with each other.

BEC has now been achieved in a variety of atomic alkali systems (Anderson et al. 1995; Davis et al. 1995; Bradley et al. 1995). The experimental techniques use a combination of laser beams and evaporative cooling to create sufficiently high phase space density for condensation to occur. In the next section we will briefly explain why alkali gases are excellent systems for studying BEC. We will then outline the conventional theoretical approach for describing such condensates, which appears to give excellent agreement with experiments. Finally, we shall discuss potential limitations of this theory (see also Proukakis et al. 1997) and outline a theoretical framework in which they can be overcome.

\section{Quantum Statistical Effects versus Interactions}

We have already argued that BEC is due to quantum statistical effects and would therefore also occur, in principle, in an ideal (i.e non-interacting) gas. However, a realistic description of experiments necessitates us taking the interactions of ultracold atoms into account. The interplay between interactions and quantum statistical effects is a rather important one for BEC, since the effects of interactions  can strongly modify the quantum statistical effects, as happens in the case of liquid Helium (see e.g. Sokol 1995). In this system, as has been known for a long time, the $\lambda$-transition (experienced at temperatures below $2.17K$) gives rise to the simultaneous existence of a `normal' fluid (just as the ones we are accustomed to) as well as a superfluid, the latter having the unique feature that it flows without viscosity. Although BEC is believed to play an important role in this behaviour, one cannot simply identify the superfluid component with a condensate (Huang 1995). The reason for this is that helium atoms interact extremely strongly with each other, making an identification of pure quantum statistical effects very hard. 

For a long time it was unclear whether one could  unambiguously observe a condensate in which the quantum statistical effects were not masked. For this reason, experimentalists have looked for ways to achieve `pure' BEC in a weakly-interacting gas. Until recently, most efforts have been focused on spin-polarised  hydrogen (Silvera \& Walraven 1980), coming very close to the BEC transition (Greytak 1995; Silvera 1995). Although BEC has not yet been achieved in this system, these efforts created very powerful tools (e.g. evaporative cooling) that where adopted in the experiments with laser-cooled trapped alkali systems, which have recently generated the first weakly-interacting condensates (Anderson et al. 1995; Davis et al. 1995; Bradley et al. 1995).

One might ask why alkali gases  are good systems for studying BEC. Experimentally, this is so, because they can be laser-cooled to temperatures in the $\mu K$ region. This is based on sub-Doppler cooling methods (Castin et al. 1995) that rely on the ground state hyperfine structure and provide the starting point for evaporative cooling (Ketterle \& van Druten 1996) that is used to cool the gas to even lower temperatures. When the temperature decreases to about a few hundred $nK$, BEC occurs in sufficiently dilute samples, which allow a clear observation of the quantum statistical effects. This is, of course, very appealing to theoreticians, as it means we can use simple weakly-interacting  Bose-gas theory and make direct predictions for the experiments. The weakly-interacting Bose-gas theories were first developed in the hope of explaining the features of liquid Helium, which they actually succeded in a qualitative rather than a quantitative way.

The temperature, $T_{c}$, at which the BEC transition takes place, does not differ greatly from the ideal gas predictions, confirming that the interactions are not strongly modifying the phenomena. This may seem perplexing at first, when one compares the translational energy of the cooled alkalis, in the $\mu K$ range, with the energy of the chemical bond between two alkali atoms which is a few orders of magnitude higher. Indeed this poses a limitation on the lifetime of the condensate, due to the three-body recombination rate (see e.g. Tiesinga et al. 1992). In this process, three-body collisions produce bound molecules and an energetic atom,  both of which escape from the trap. However, on the short timescales in which experiments are performed (a few seconds for current condensates), one can  consider the atoms as weakly-interacting, with the dominant process being that of elastic binary collisions. To understand how this is possible, we need to remember that the atoms are at such low temperatures that their de Broglie wavelengths are huge compared to the range of the interactions. This means that the net effect of the interactions  on the atomic properties at large distances can be replaced by a simple shift in the atomic de Broglie wave. Hence, we can model the ultracold alkali interactions as hard-sphere interactions, where the radius of the hard sphere is exactly this above shift in the de Broglie wave. This shift, known as the s-wave scattering length is typically of the order of a few tens of Bohr radii, i.e. a few nm for the heavier alkalis. Experimental measurements of scattering lengths have been made for most alkali systems (see e.g. Tiesinga et al. 1996). Since the scattering length is so small in comparison to the de Broglie wavelength of the ultracold atoms, the gas is effectively dilute. This means the effect of interatomic interactions on the gas remains modest and ensures that the BEC transition is not strongly modified by them.

The interactions do, however, become more important in the condensed phase, as more and more atoms are preferentially scattered into the ground state of the system. For the homogeneous gas, this results in a peak in the momentum distribution, whereas, in the case of a trap, to a peak in the density of atoms. This increased density implies that the interactions between the condensed atoms will strongly affect the properties of the condensate as well as its interaction with the non-condensed particles. Nonetheless, the range of the interactions remains extremely small compared to the de Broglie wavelength or the interparticle spacing, so that we might expect to be able to treat the ultracold collisions using a scattering length model. Mathematically this means that we can simplify the spatially-dependent interaction potential between two ultracold alkali atoms into that of a zero-range delta-function potential (see e.g. Huang 1987)
\begin{equation}
V({\bf r}) = U_{0} \delta({\bf r})
\end{equation}
Here $U_{0}$ represents the effective interaction strength for zero-range interactions, which is related to the s-wave scattering length, a, by $U_{0} = \frac{4 \pi \hbar^{2} a}{m}$ where m is the atomic mass. Approximating the potential in the form (1) enables us --- as will become clear below --- to make a direct link between theory and experiment. The scattering lengths used in calculations have been found in other spectroscopic determinations. Approaches based on this potential, along with the Gross-Pitaevskii equation we shall now discuss, have given impressive results. There are, however, as we shall see, some serious conceptual difficulties with the microscopic basis of the calculation.

\section{The Gross-Pitaevskii Equation for the Condensate Wavefunction}

In order to describe the evolution of a trapped condensate, we shall use mean-field theory. Consider the motion of an atom in an interacting Bose-condensed assembly. Since the atoms are Bose-condensed, that is they exist in the same quantum state, they can be thought of as acting coherently on our single atom. Thus, our atom is not aware of the individual behaviour of each atom in the assembly and behaves dominantly as if it were moving through the condensate mean field. 

Let us consider the time-dependence of the wavefunction $\Phi({\bf r}, t)$ of an atom in the condensate. The use of mean-field theory as described above, along with the interaction potential of equation (1) give the following equation for the evolution of this single-particle wavefunction 
\begin{equation}
i \hbar \frac{\partial \Phi({\bf r} ,t)}{\partial t} = \left( - \frac{\hbar^{2} \nabla_{{\bf r}}^{2}}{2m} + V_{trap}({\bf r}) \right) \Phi({\bf r}, t ) + NU_{0}  | \Phi({\bf r} ,t) |^{2}  \Phi({\bf r} ,t) 
\end{equation}
The first term in the above equation describes the free evolution of the atom in a trapping potential $ V_{trap}({\bf r})$. The second term expresses the coherent action of all other condensate atoms on the particular atom under consideration, i.e. the effect of the condensate (or Hartree) mean-field, and $N$ corresponds to the number of atoms in the condensate. Now let us assume that our system is cooled to temperatures so close to absolute zero and  that effectively all atoms in the assembly are condensed. In this case, all atoms in the condensate will be described by this same wavefunction, which we can term the wavefunction for the condensate. This last step of assigning the single-atom wavefunction to describe the coherent effect of the whole assembly is, however, not exact. The reason for this is that even at $T = 0$, the interactions between the atoms will cause a depletion of the condensate (Lifshitz \& Pitaevskii 1980). 

The depletion of the condensate can be understood in both a particle and a quantum field theory point of view. Let us look at the particle point of view first. The idea of assigning a single wavefunction for all the particles means that the probability of finding a particle in a given region is independent of the position of any other particle in the assembly. This approximation clearly has to fail in the region where two atoms approach within the range of their mutual molecular interactions. One would therefore expect a depletion of the pure uncorrelated condensate wavefunction that would depend on the mean number of atoms that find themselves within the range of influence of a neighbouring atom. This means we might expect, and we do indeed find, that the depletion depends on $n a^{3}$ (Lifshitz \& Pitaevskii 1980), where $n$ corresponds to the condensate number density. 

In the case of liquid Helium this depletion is so large (around 90\%), that a mean-field approach cannot be used to obtain quantitative results. For the case of the alkali condensates produced to date, we can thus use typical values (for heavy alkalis $a \sim 5 nm $ and $n \sim 10^{14} cm^{-3} $) to estimate the depletion to lie in the 1\% range. Actually, it is possible to calculate this more precisely. A recent theoretical study (Hutchinson et al. 1997) of the $T=0$ depletion in the case of 2000 Rb atoms\footnote{This study corresponds roughly to the first experimental observation of BEC at JILA (Anderson et al. 1995), although we stress that the JILA TOP trap was actually anisotropic.} in an isotropic trap has shown this depletion to be of the order of $0.5\%$, a truly negligible figure. Thus, we are justified in identifying the wavefunction of a single atom with that of the whole condensed assembly, which enables us to study the evolution of the condensate mean-field. We should point out that this can only be valid to order $\frac{1}{N}$, $N$ being the total number of particles in the condensate. 

Equation (2) is known as the Gross-Pitaevskii or nonlinear Schr\"{o}dinger equation (Ginzburg \& Pitaevskii 1958; Gross 1963). It has been used very widely in determing the properties of Bose-condensates at near-zero temperatures and appears to provide excellent predictions for relevant experimental observations (see Ruprecht et al. 1995 and 1996; Edwards et al. 1996a-c; Dodd 1996 a,b; Jin et al. 1996; Mewes et al. 1996; Stringari 1996, etc.). Nonetheless, examination of its microscopic basis (Proukakis and Burnett 1996a, Proukakis et al. 1997) casts doubts on its validity in certain regimes, as we shall explain shortly. In the quantum field theory picture of condensation, the Gross-Pitaevskii equation can be thought of as the equation expressing the evolution of the mean value of the field operator for the bosonic atom field. The fluctuations from this mean value then represent fluctuations in the mean field.

The finite depletion mentioned above implies that even at $T = 0$, there are still some atoms which are not in the condensate, due to the interactions between them. In all theoretical treatments, we assume that there is a well-defined condensate mean-field and that these excited atoms, or excitations, move in the presence of the condensate. This technique of assuming a well-defined (condensate) mean-field and splitting off the remaining effects due to elementary excitations around it is common to the treatment of many systems with a broken symmetry (e.g. in the case of BEC, the condensate is a broken symmetry state).

Naturally, we expect the presence of a condensate mean-field to affect the spectrum of the elementary excitations of our bosonic gas. The reason for this is that the atoms are not interacting with other individual atoms, but are moving dominantly in the presence of a coherent condensate field. As expected, the deviation from the non-interacting spectrum becomes more pronounced, the larger the number of particles added to the condensate. The notion of an excited atom interacting with condensed atoms leads quite naturally to the idea of a quasiparticle. One can see the significance of the change in the spectrum by looking at the quasiparticle dispersion relation in the case of a homogeneous (i.e. free) condensate, namely (Lifshitz and Pitaevskii 1980)
\begin{equation}
\hbar \omega_{k} = \sqrt{\epsilon_{k}^{2} + 2 n U_{0} \epsilon_{k}}
\end{equation}
where $\epsilon_{k} = \frac{\hbar^{2} k^{2}}{2m}$ corresponds to the non-interacting energy spectrum and $n$ to the condensate number density. Clearly, in the case of small condensate densities, the contribution of the second term is rather small and a quasiparticle identically reduces to a particle in the absence of condensation. The same is true for large momenta $k$, where the first term dominates; the physical justification here is that these atoms are moving too fast to be influenced by the condensate mean field. However, for small $k$, i.e. slower speeds, the second term becomes increasingly important, resulting in  a phonon-like spectrum for small $k$. This can be physically explained by the fact that all atoms start in the same state (the condensate), so that any attempt to disturb a single one of them, results in a coherent disturbance of all of them.

In the inhomogeneous case, the excitation spectrum can be determined by means of the Gross-Pitaevskii equation by linearising around the ground state $\phi_{g}$ of the condensate, according to (Ruprecht et al. 1996)
\begin{equation}
\Phi({\bf r}) = \phi_{g}({\bf r}) + u_{\lambda}({\bf r}) e^{ -i \omega_{\lambda} t} +  v_{\lambda}({\bf r}) e^{ + i \omega_{\lambda} t}
\end{equation}
In this method, we treat excitations out of the condensate  as ripples of frequency $\omega_{\lambda}$ on top of it. The spatially-dependent  coefficients $u_{\lambda}({\bf r})$  and $v_{\lambda}({\bf r})$ express the amount of quasiparticle dressing and depend on the strength of the interactions. In the limit of excitation energies much larger than $n U_{0}$, $v_{\lambda}({\bf r}) \rightarrow 0$ and $u_{\lambda}({\bf r})$ goes over to being the  eigenfunction of the free simple harmonic oscillator.

This linearisation technique, well-known from response theory, is equivalent to the more formal one of diagonalising the hamiltonian of the system by means of a transformation to quasiparticles (Edwards et al. 1996c). The set of equations we thus obtain from these methods determine the shapes and frequencies of the excitations of the condensate. Simulations done for the Rb condensates in the JILA TOP trap using this method have given predictions which agree with the experimental measurements to within 5\% (Edwards et al. 1996b), an extraordinary feat given the simplicity of this approach. Furthermore, we wish to point out that the predictions agree very well with predictions based on a hydrodynamic approach (Stringari 1996) in the appropriate large condensate limit. This excellent agreement has generated a lot of confidence in the use of mean field theory, and its possible extension to finite temperatures.

So far we have indicated that the Gross-Pitaevskii equation gives excellent predictions for the excitations of Bose-Einstein condensates. However, the theory is extremely simple and can only be used at temperatures very close to absolute zero. Furthermore, damping processeses which have now been observed in experiments (Jin et al. 1997) are not included in it. Recent work has focused on the development of mean-field theory valid at finite temperatures, and we shall discuss how we have gone about deriving such a theory.

\section{Microscopic Mean Field Theory}

The Gross-Pitaevskii equation has been widely applied in a variety of different situations. Nevertheless, the discussion of its validity regime in the literature is rather limited. The limitation to $T = 0$ is clear, but there are other issues that need attention. In order to investigate other limitations of this theory, we have extended mean field theory by considering fluctuations around the pure condensate mean field and their temporal evolution for trapped atoms (Proukakis \& Burnett 1996a)\footnote{We would like to point out that some effects of these fluctuations have also been considered by Stoof (1992) and Biljsma and Stoof (1997), who came to similar conclusions to ours.}.

The conventional approach of equations (1) and (2) is essentially a phenomenological one. In our treatment, however, we have looked at the system on the microscopic scale and examined the detailed effect of collisions taking place in the system. In this way, we have developed a mean field theory for both zero and finite temperatures (Proukakis et al. 1997). Let us initially consider our microscopic approach at $T = 0$, which enables us to examine the limitations of the Gross-Pitaevskii equation.

\subsection{$T = 0$ Mean Field Theory and the T-matrix}

In a dilute Bose gas near $T = 0$, most of the atoms are in the condensate. It is thus reasonable to focus our theoretical treatment on the interactions between pairs of condensed atoms. The interaction potential will promote pairs of atoms out of the condensate, and, once excited, they may re-interact with each other. The result of this second `single-vertex' interaction varies: the atoms can either fall back into the condensate, or make a transition to different non-condensate states. This `single-vertex' interaction can be repeated an arbitrary number of times, before the collision ends, with both atoms occupying condensed states once again. A completed collision, or scattering, in which both initial and final collisional states are condensed ones, is therefore made up of an arbitrary number of intermediate excited-state pairs, interacting via a `single-vertex' interaction $V({\bf r})$.

A microscopic treatment which includes fluctuations around the pure condensate mean field can be shown to include the effect of all repeated `single-vertex' interactions, thus accurately describing the scattering of two condensed atoms in a dilute trapped condensate at $T =0$. In other words, the interaction potential between two condensate atoms gets `renormalised' to the two-body T-matrix, which accounts for the full effect of a completed collision (in vacuum).

We would like to stress, however, that it is precisely the fluctuations around the condensate mean field which cause this `renormalisation' of the interaction potential from the `single-vertex' $V({\bf r})$ to the T-matrix. The effect of such fluctuations are assumed to be negligible in the simplest conventional derivation of the Gross-Pitaevskii equation. In this approach to the Gross-Pitaevskii equation, one takes in effect a `single-vertex' interaction with both initial and final states being in the condensate (i.e the most trivial collisional process that can occur). Such an approach clearly cannot treat the evolution of the collision via intermediate states, and we shall argue that this is where the most interesting collisional dynamics occur. To overcome this problem, one conveniently makes the assumption, flawed in our opinion, that this `single-vertex' interaction in the Gross-Pitaevskii equation treats fully the condensate-condensate scattering process. The reason why this last step is flawed, is that the only way to upgrade the `single-vertex' interaction to a completed collision (i.e. $V({\bf r}) \rightarrow$ T-matrix) is by suitable treatment of the fluctuations around the condensate mean field, and yet the conventional derivation of the Gross-Pitaevskii equation disregards these very fluctuations as negligible (e.g. Nozi\`{e}res and Pines 1990).

The intermediate collisional states treated by these fluctuations will, however, be modified due to the fact that the atoms are not colliding in vacuum, but in the presence of a condensate, and such modifications can only be taken into account by a microscopic treatment of the fluctuations. The presence of such additional intermediate processes invalidates the claim that any corrections on the Gross-Pitaevskii equation are purely those due to deviations from the dilute gas limit.

We want to stress that it is only the completed atom-atom scattering process in vacuum which can be accurately described by the two-body T-matrix and its approximate replacement  by a $\delta$-function. The validity of such an approximation has been discussed elsewhere (Proukakis et al. 1997; Huang 1987). If it were true that such a condensate-condensate collision could be accurately represented in vacuum, then our treatment to this point would merely put the Gross-Pitaevskii equation on firm microscopic basis\footnote{This would be somewhat similar to the work of Beliaev (1958) and Popov (1987).}. In what follows, we shall deal with the effect of the atomic medium on the binary collisions and explain why these effects cannot be ignored.

\subsection{Finite Temperature Mean Field Theory}

\subsubsection{The Many-Body T-matrix}

We have discussed earlier how the scattering of two condensate atoms proceeds via a sequence of excited-state pairs. In the $T = 0 $ limit, we can, to a good approximation, treat those intermediate states as empty. When, however, the temperature of the atomic assembly is increased, atoms will be thermally excited to non-condensate states. We have already mentioned that the nature of bosonic statistics implies enhanced scattering into occupied states. Thus the condensate-condensate scattering amplitude will also be modified due to this occupation of excited states through which the scattering proceeds. This means that the effective interaction, mentioned in the previous section, is now represented by the many-body T-matrix which explicitly takes account of the occupation of the intermediate collisional states\footnote{A more detailed account of the two-body versus the many-body T-matrix can be found in the review article by Stoof et al (1996).}. 

In the case of repulsive interactions, the modification from the two-body to the many-body T-matrix in three-dimensions is modest, except close to the transition temperature. Close to $T_{c}$, Bijlsma and Stoof (1996) have shown, however, that it causes a profound renormalisation of the T-matrix. In the case of attractive interactions, the  many-body T-matrix, can diverge, signalling the onset of a BCS transition in the gas (Stoof 1994). In both of these cases, diluteness does not prevent the effects setting in.

\subsubsection{The Mean Field of Thermally Excited States}

The population in excited states also produces a mean field which will modify the evolution of condensed atoms. We would thus expect an additional term in equation (2), depending on both the amount of excited state population, as well as the condensate mean field.

In order to obtain a consistent theory, this condensate-excited state interaction should be described by the many-body T-matrix. However, this can only be brought into the equation (Proukakis et al. 1997), by considering even more complex fluctuations around the condensate mean field (corresponding to 3-body interactions). This implies that the use of finite temperature mean field theory for condensate evolution can only be consistent when one goes beyond the conventional (Hartree-Fock-Bogoliubov or HFB\footnote{For a description of the conventional Hartree-Fock-Bogoliubov theory, see e.g. Kobe 1968; D\"{o}rre et al. 1979; Huse and Siggia 1982.}) approximation. We believe this inconsistency also affects the $T=0$ limit of the Gross-Pitaevskii equation, because of the finite (but small) population in excited states.

Let us return to our discussion of the importance of the corrections on the Gross-Pitaevskii equation due to the fluctuations around the condensate mean field. In the chemical potential of the homogeneous gas, the correction is well-known to be
\begin{equation}
\mu = n U_{0} \rightarrow n U_{0} \left(1 + \frac{32}{3} \sqrt{ \frac{n a^{3}}{\pi}} \right)
\end{equation}
However, our treatment shows that in order to obtain precisely this correction, we must not only take account of the effective condensate-condensate scattering\footnote{In considering the condensate-condensate scattering process , we point out that the `renormalisation' of $V({\bf r})$ to the T-matrix is implicit in the HFB treatment.}, but we must further include the effect of the few non-condensate states present near $T = 0$. The correct inclusion of this effect necessitates the consideration of fluctuations beyond the conventional HFB treatment. This suggests that the HFB approach is, strictly speaking, not sufficient for the description of the condensate, even at $T=0$ (although in the case of the weakly-interacting systems under consideration, this limitation is really an academic one). At this point, the reader may well be wondering whether the conventional approach can be made consistent, by means of the microscopic method we have developed.

In fact, when one extends the treatment beyond HFB, one finds that the extra fluctuations play a two-fold role: firstly, they `renormalise' the condensate-excited states scattering potential $V({\bf r})$ to an effective T-matrix interaction, which has already been used in obtaining (5). However, the beyond-HFB approach also introduces new physical processes that dress the binary interactions due to the mean field (Proukakis et al. 1997). This dressing is additional to that included in the HFB treatment. Hence, the statement that the fluctuations ignored in the conventional derivation of then Gross-Pitaevskii equation, merely result in corrections of the form (5) is only correct in the limit when the effect of the mean field\footnote{We remind the reader that by mean field we are referring to the mean field of both condensed, as well as excited atoms.} on the intermediate collisional states can be ignored.

\subsection{Dressing in the Intermediate Collisional States}

To obtain a consistent microscopic approach to the dilute Bose gas, we must therefore take into account the modification to the collision dynamics due to the presence of the mean field. Such a mean field `dresses' the intermediate states of a collision, and, in the case of an infinte gas near $T=0$, as is well-known from other contexts, causes a divergence in the effective interaction (termed the `dressed' or quasiparticle many-body T-matrix), when treated in perturbation theory (Nepomnyashchii, Y. A. \& Nepomnyashchii, A. A. 1978). A more sophisticated analysis shows that this apparent divergence leads to the effective interaction strength in the nonlinear Schr\"{o}dinger equation for a homogeneous gas going to zero (Griffin 1993), which indicates that such a theory should be treated with caution. 

In the case of inhomogeneous systems, the many-body T-matrix cannot diverge. As long as the energy spacing of the trap is larger than $n U_{0}$, the effects of the condensate mean field on any of the states can be ignored. We can then state one of the rigorous conditions for the Gross-Pitaevskii equation to hold, namely that, for positive scattering lengths (Proukakis et al. 1997)
\begin{equation}
n U_{0} \ll \hbar \omega
\end{equation}

In our opinion, the finite temperature calculations to date based on ignoring these rather important issues, may encounter problems. One of the ways out of this problem is to reformulate mean-field theory for inhomogeneous systems in the style of Popov's theory (Popov 1987) for the homogeneous Bose gas (i.e in terms of an effective condensate density and phase). Part of this work has been achieved by Ilinsky and Stepanenko (1997), but much remains to be investigated. It is quite likely that a mean field theory that produces results consistent with experiment can be produced by a judicious choice of approximations. However, it is still uncertain whether such a theory can be put on a fully microscopic basis.

\section{Conclusions}

In this paper we have given an account of the physical processes that have to be considered in the microscopic derivation of mean-field theory. They lead us to be wary in the application of the present versions of finite temperature mean field theory in the analysis of trapped Bose-Einstein condensed gases. Time will tell how effective they really are in making quantitative predictions. Our account should certainly not prevent people from making predictions, but should, we believe, alert them to the potential troubles ahead.

\begin{acknowledgments}

There are a number of people we would like to thank, particularly our collaborators in Oxford, the University of Utrecht (the Netherlands) and the National Institute of Standards and Technology in the U.S. The research of N. P. Proukakis is supported by the Alexander S. Onassis Public Benefit Foundation (Greece) and K. Burnett would like to acknowledge financial support from the U.K. EPSRC.

\end{acknowledgments}

\end{document}